\documentclass[10pt,letterpaper]{article}
\usepackage{opex3}
\usepackage{cite}
\usepackage{dcolumn}   
\usepackage{bm}        
\usepackage{amssymb}   
\usepackage{verbatim}  
\usepackage{epstopdf}

\newcommand{\beq}{\begin{equation}}
\newcommand{\eeq}{\end{equation}}
\newcommand{\bea}{\begin{eqnarray}}
\newcommand{\eea}{\end{eqnarray}}

\newcommand{\eqnjp}[1]{{(\ref{#1})}}
\newcommand{\commentout}[1]{{}}

\newcommand{\bE}{{\bf E}}
\newcommand{\bor}{{\bf r}}

\newcommand{\dip}{{\cal D}}
\newcommand{\G}{{\sf G}}
\newcommand{\bd}{{\bf d}}

\begin{document}

\title{Light propagation beyond the mean-field theory of standard optics}

\author{Juha Javanainen$^{1,*} $ and Janne Ruostekoski$^{2}$}

\address{$^1$Department of Physics, University of Connecticut, Storrs, Connecticut 06269-3046, USA\\
$^2$Mathematical Sciences, University of Southampton,
Southampton, SO17 1BJ, UK}

\email{$^*$jj@phys.uconn.edu} 



\begin{abstract}
With ready access to massive computer clusters we may now study light propagation in a dense cold atomic gas by means of basically exact numerical simulations. We report on a direct comparison between traditional optics, that is, electrodynamics of a polarizable medium, and numerical simulations in an elementary  problem of light propagating through a slab of matter. The standard optics fails already at quite low atom densities, and the failure becomes dramatic when the average interatomic separation is reduced to around $k^{-1}$, where $k$ is the wave number of resonant light. The difference between the two solutions originates from correlations between the atoms induced by light-mediated dipole-dipole interactions.
\end{abstract}


%
%



\section{Introduction}

Traditional analysis of light propagation in a medium is based on more than a century-old  electrodynamics of a polarizable medium (EDPM) and its consequences such as standard optics \cite{JAC99,BOR99}. The usual EDPM, however, is an effective-medium mean-field theory (MFT): While atomic ensembles are composed of essentially point-like dipolar emitters, EDPM assumes that each atom interacts with the average behavior of the neighboring atoms. MFT models are commonplace in condensed-matter physics, and the usual
observation is that they become inaccurate when the particles become correlated due to strong interparticle interactions. Light incident on an optical medium induces resonant dipole-dipole interactions between the atoms. The question we raise is:
can the dipole-dipole interactions in light propagation be strong enough that the standard optics
fails?

We present essentially exact large-scale simulations of light propagation in a cold (laser-cooled or evaporatively cooled) and dense medium of discrete atoms. In fact, the small gas samples available in the laboratories \cite{BEN10,BAL13,KEA12,PEL14,KEM14} have recently made it possible to simulate atom-light systems numerically \cite{PEL14,JAV99LQD,PIN04,Jenkinsoptlattice,CHO12,JEN12a,Castin2013,Skipetrov14,JAV14} approaching experimentally realistic conditions \cite{JAV14,PEL14}.
We compare the numerical solutions side by side with the predictions of EDPM in an elementary optics problem, propagation of light through a slab of matter, and at high gas density demonstrate a dramatic failure of the standard optics. We also find deviations from EDPM at surprising low atom densities, and with the light far off atomic resonance. The new phenomenology is currently becoming accessible to dedicated experiments, and may even occur inadvertently where cold and dense atomic gases are studied experimentally.

We consider the limit of low light intensity and a low-temperature sample modelled with stationary atoms, so that the exact solution is obtained by purely \emph{classical} microscopic electrodynamics~\cite{JAV97,JAV99LQD}.
Quantum effects therefore cannot explain away the discrepancies
between the exact solution and the predictions of EDPM, but the failure of EDPM indicates light-induced correlations between the atoms~\cite{MOR95,JAV97,Jenkinsoptlattice,JAV14}.
Our simulations suggest that the emergence of macroscopic classical electromagnetism from microscopic principles may be more complex than previously thought, and may involve an as-of-yet undiscovered element of physics.

\section{Standard optics versus exact numerical solution}

We take the elementary-textbook problem of light propagating perpendicularly through a slab of a dielectric medium of thickness $h$. The MFT solution for the transmitted field in such a geometry is a simple exercise in optics. For comparison with the standard optics, we solve the quantum field theory of light  propagation~\cite{JAV97} for stationary atoms in the weak excitation limit essentially exactly\cite{JAV97,JAV99LQD}. The method is based on classical  electrodynamics simulations \cite{JAV99LQD,PIN04,Jenkinsoptlattice,CHO12,JEN12a,Castin2013} and
stochastic Monte-Carlo sampling of the positions of the atomic dipoles. The primary computed quantity is the optical thickness $D$ related to the power transmission coefficient $T$ by $D=-\ln T$. The relevant parameters are the wavenumber of the light $k$, the thickness and density of the sample $h$ and $\rho$, HWHM resonance linewidth of the atoms $\gamma$, and the detuning of the light from atomic resonance $\Delta$. The Lorentz-Lorenz (LL) shift $\Delta_{LL}= -2\pi\gamma\rho k^{-3}$ affords a convenient scaling to express density as a frequency. In numerical simulations we use a circular disk of area $A$ instead of an infinite slab, which results in a truncation error. We cite the area and an estimate for the error where appropriate. Appendices A and B give more technical details.

\begin{figure}
\vspace{-12pt}
\center\includegraphics[width=0.65\columnwidth]{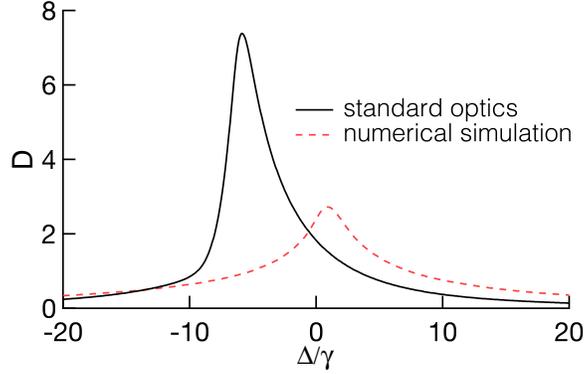}
\vspace{-12pt}
\caption{Optical thickness  (depth, density) of a slab of matter as a function of light-atom detuning from both standard optics and from an essentially exact numerical simulation for stationary strongly interacting atomic dipoles. The results are for the sample density $\rho = k^3$ and slab thickness $h=k^{-1}$. The truncation error in the numerical computations due to the finite area $A = 1024\,k^{-2}$ of the disk-shape sample, about 5\%, is irrelevant for a qualitative comparison. The conspicuous difference between the two curves demonstrates the breakdown of traditional optics.}
\label{SLABTRANSMISSION}
\end{figure}

Figure~\ref{SLABTRANSMISSION} illustrates a dramatic breakdown of traditional optics and the corresponding continuous-medium MFT for near-resonant light at the atom density  $\rho=k^3$. The sample thickness $h=k^{-1}$  is similar to the sample dimensions in some of the most recent experiments \cite{PEL14,KEA12}.

Next we wish to study the emergence of optical phenomenology beyond the MFT in two limits when the light-mediated atom-atom interactions are expected to be weak: When the driving light is tuned far off resonance, and when the distance between the atoms increases with decreasing atom density.

\noindent \emph{Light-induced correlations in the far off-resonant limit}.
For the present purposes we define the absorption coefficient ${\cal A}$ in terms of the optical thickness or ``depth'' $D$ and transmission coefficient $T$  by ${\cal A}=1-T=1-e^{-D}$. For the MFT, Eqs.~\eqnjp{ET} and~\eqnjp{NSQR} in Appendix~\ref{SEM}, an expansion in $1/\Delta$ appropriate in the limit of a large detuning $\Delta$ gives
\beq
{\cal A}_M = \frac{6\pi\rho h  \gamma^2}{k^2 \Delta^2}\left[1+\frac{3|\Delta_{\rm LL}|}{8hk\gamma}(1-\cos 2hk) \right] + {\cal O}\left(\frac{1}{\Delta^3}\right).
\eeq
The factor in front is the absorption coefficient of independent atoms far-off resonance when the cross section for photon scattering $\simeq{6\pi \gamma^2}/({k^2 \Delta^2})$ is asymptotically small. Maybe surprisingly, even in MFT and far-off resonance, the sample does not have to behave like a collection of independent radiators. Here we see remnants of the etalon effect due to the reflections of light at the faces of the slab expressed in terms of the LL shift that is proportional to atom density.

\begin{figure}
\vspace{-12pt}
\center\includegraphics[width=0.65\columnwidth]{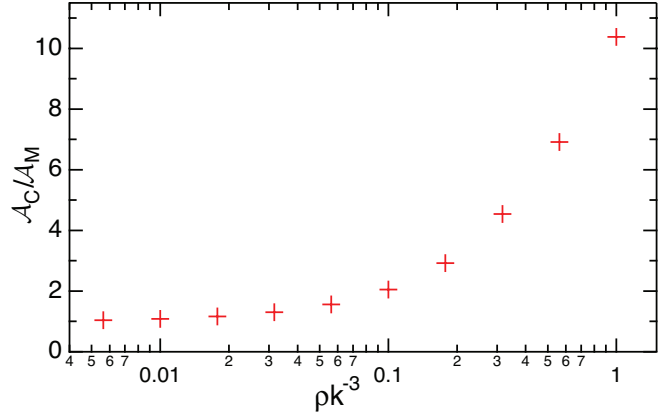}
\vspace{-6pt}
\caption{Ratio of the absorption coefficient from numerical computations, ${\cal A}_C$, and from MFT, ${\cal A}_M$, as a function of the density of the slab with the thickness $h=\pi/k$. The positive detunings are chosen in such a way that the MFT gives ${\cal A}_M =0.01$ for each density, an optically thin sample. The mean free path for light scattering for independently radiating atoms  would therefore be approximately $100$ times the thickness of the sample for all data points. Nevertheless, there are up to factor-of-ten deviations from the MFT. The numbers of atoms, obviously integers, are chosen so that for a given density the area of the disk is as close to $A=2000\,k^{-2}$ as possible, whereupon the truncation errors in the absorption coefficients ${\cal A}_C$ are at most a few per cent.}
\label{SCATTERING}
\end{figure}

In comparisons with numerical simulations, we fix the thickness of the slab arbitrarily at $h=\pi/k$. For various values of density $\rho$ we then find from the MFT exactly, numerically, the corresponding positive detuning $\Delta(\rho)$ for which the absorption coefficient is ${\cal A}_M=0.01$. The slab should therefore be optically thin. To corroborate cooperative light scattering, we compute  the absorption coefficients ${\cal A}_C$ from our beyond-MFT classical electrodynamics simulations for the same densities $\rho$ and detunings $\Delta(\rho)$. In Fig.~2 we plot the ratio ${\cal A}_C/{\cal A}_M$, the cooperative enhancement of absorption, for a number of sample densities. It exceeds two already for a sample as dilute as $\rho k^{-3}=0.1$, and is more than ten at $\rho k^{-3} = 1$.

This example clearly shows that the deviations from the MFT are governed by the parameter~$\rho k^{-3}$, not optical thickness per se. Our interpretation is that $\rho k^{-3}$ is the dimensionless density parameter for cooperative effects. There is also another dimensionless density parameter present, on-resonance optical thickness $6\pi\rho hk^{-2}$, which is a MFT concept. This distinction is typically not made in the literature on cooperative effects. Unfortunately, as in dense samples with $\rho k^{-3}\sim1$ the numerical effort limits the range of sample thickness available for simulations to at most a few $k^{-1}$, we are yet to find a practical way to vary the two parameters independently over wide ranges. Therefore we cannot make a clean distinction between them either.

\noindent\emph{Light-induced correlations in the limit of low atom density}.
Next we expand the optical depth derived from the MFT expressions~\eqnjp{ET} and \eqnjp{NSQR} in density $\rho$, and choose the parameters $K_0$ and $K_1$ independent of $\rho$ in the analytical model $D=\rho/(K_0 + K_1\rho)$ in such a way that to second order in $\rho$ we have the same expansion. The quantity  $D$ then presents the usual resonance lineshape as a function of the detuning $\Delta$, with the resonance shifted from $\Delta=0$ by $s$:
\bea
s&=& \Delta_{\rm LL} +\hbox{$\frac{3}{4}$} |\Delta_{\rm LL}|\left( 1-\frac{\sin 2hk}{2hk}\right).
\label{COLLAMB}
\eea
We find the same shift also numerically when we fit a Lorentzian with a variable resonance position  to the MFT absorption line shape, and go to the limit of low sample density.
The result shows the LL shift, plus etalon effects. The etalon contribution equals the ``cooperative Lamb shift'' originally introduced by Friedberg, Hartmann, and Manassah \cite{FRI73}. The theoretical result~\eqnjp{COLLAMB} was recently tested experimentally, albeit in an inhomogeneously broadened gas~\cite{KEA12}.

\begin{figure}
\vspace{-12pt}
\center\includegraphics[width=0.7\columnwidth]{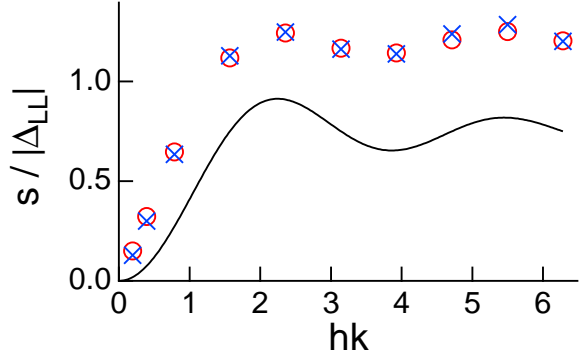}
\vspace{-12pt}

\caption{Shift of the resonance $s$, scaled to the absolute value of the LL shift $|\Delta_{\rm LL}|$, as a function of the sample thickness $h$, for two densities $\rho=0.01\,k^3$ (circles) and $\rho=0.005\,k^3$ (crosses). Also shown is the prediction~\eqnjp{COLLAMB}, shifted up by $|\Delta_{\rm LL}|$ for easier comparison (solid line). The largest optical thickness in the raw data is $D\simeq1$, so for all parameter values in this figure the sample is still fairly transparent. The truncation errors in the data points are at most on the order of one per cent.}
\label{SHIFTS}
\end{figure}

To compare the MFT shift with the numerical simulations we pick two densities $\rho = 0.01\,k^3$ and $\rho = 0.005\,k^3$. We compute the optical thickness as a function of detuning, and fit the results to a Lorentzian with a variable width and shift $s$.   In Fig.~\ref{SHIFTS}  we show  the shifts as a function of the sample thickness in units of the LL shift for each density. We also display the prediction~(\ref{COLLAMB}), omitting the leading LL shift that would move the curve down by one unit.

When the thickness of the sample decreases, in the end the physics must become two-dimensional~\cite{CHO12} and area density, not density, is the relevant parameter. Besides, since we keep the density constant, in the limit $h\rightarrow0$ the area density $h\rho$ vanishes. That is why \cite{JAV14} the shift vanishes in the limit $h\rightarrow0$.  Starting from about $h=k^{-1}$ an oscillatory dependence on sample thickness is found that is qualitatively similar to the result~\eqnjp{COLLAMB}, with an approximately constant difference from the prediction~\eqnjp{COLLAMB}. We take this to indicate that three-dimensional  bulk-sample physics sets on at the thickness of about $h\sim k^{-1}$. Without the scaling by $|\Delta_{\rm LL}|$ the difference from~(\ref{COLLAMB}) would be proportional to sample density \cite{KEA12}. This difference is a beyond-MFT effect.

At low sample densities one expects a shift of a resonance proportional to $\rho k^{-3}$ from dimensional analysis, but dimensional analysis alone does not provide the multiplicative constant. The LL shift amounts to a specific prediction for the constant. The usual local-field corrections and the ensuing Lorentz-Lorenz and Clausius-Mossotti formulas depend on this particular value of the constant. However, it is well known that the atom-atom correlations ignored in MFT produce corrections in the susceptibility of the gas that are of the same order in density as the LL shift~\cite{MOR95,RUO99}, so  one has to wonder if the LL shift is all there is to it. In fact, none of our simulations here or in~\cite{JAV14} agree with the LL shift. The enduring success of the textbook local-field corrections remains a mystery to us.

\section{Criterion for cooperative atom response}

Our follow-up agenda is to set up a framework to think about cooperative effects in light propagation. In common parlance cooperativity entails that each emitter (henceforth, atom) has the radiation from the other atoms fall on it, which in turn alters the field that the atom emits. Unfortunately, this definition implies that even nearly {\em all of\/} EDPM is about cooperative effects.
We propose that a meaningful notion of cooperativity should come with the condition that cooperative effects reflect strongly correlated behavior of the radiators.  This leads to a practical criterion that the response of a sample to light should not be deemed cooperative if it can be explained using  EDPM.

In three dimensions it is surprisingly difficult to solve Maxwell's equations for a continuous polarizable medium even numerically. For small atomic samples such as those in~\cite{PEL14} it is actually easier to simulate the entire strongly interacting theory. In the recent experiments striving to demonstrate cooperative behavior in the light scattered from a dense cold gas \cite{BEN10,BAL13,PEL14,KEM14} standard EDPM has therefore not been ruled out as the cause of the observations. However, in our slab example the exact EDPM solution is on hand for comparison with materially exact numerical simulations. We not only may study the emergence of beyond-standard optics response to light quantitatively, but also make the observation that the  ``cooperative Lamb shift'' in a slab sample \cite{FRI73,ROH10,SCU10,KEA12} arises from elementary optics and is not cooperative by our criterion.

\section{Discussion and conclusions}

The EDPM is a MFT that treats the surroundings of each radiator in the average sense, as if the other atoms made a continuous polarizable medium. The MFT may be implemented technically~\cite{JAV14} by factoring a two-point correlation function of polarization and ground-state atom density as the product of polarization and density, as in ${\bf P}_2(\bor_1;\bor_2)={\bf P}_1(\bor_1)\rho(\bor_2)$. This approximation amounts to neglecting position-dependent light-induced correlations between the atoms: In the presence of the dipole-dipole interactions the dipole moment of an atom obviously depends on precisely where the nearby atoms are. Our simulations go beyond the MFT in that they include all such correlations exactly. The difference between the numerical simulations and the MFT results therefore indicates the presence of the correlations.

Subwavelength thickness inhomogeneously broadened (hot) atomic samples bounded by glass walls \cite{KEA12} and ellipsoid-shape small homogeneously broadened (cold) trapped atom samples \cite{BAL13,BIE13,KEA12,PEL14,KEM14} have been used to study  light-matter interactions, but the combination of wall-less confinement and homogeneous broadening is yet to be achieved experimentally for a slab. At the moment our analysis constitutes a thought experiment, but we surmise that the lesson is generic: As the dipole-dipole interactions grow in significance, for instance with increasing atom density, so grow the deviations of the response of the atoms from standard electrodynamics; and by the time the density reaches $\rho \sim k^3$ the difference can be dramatic.

One may wonder if diffuse scattering plays a role in the observed discrepancy, but this appears unlikely in our examples. For instance, it is difficult to see how diffusion could be the reason for the factor-of-two deviations from the MFT in Fig.~\ref{SCATTERING} when the estimated mean free path of photons in the medium is a hundred times the thickness of the sample.

In short, there is more than we have learned from the venerable textbooks to light propagation  in cold, dense gases that laser cooling and evaporative cooling now provide for the experiments.

\appendix
\section{Standard electrodynamics for a slab of atoms}\label{SEM}
\begin{figure}
\center\includegraphics[width=0.5\columnwidth]{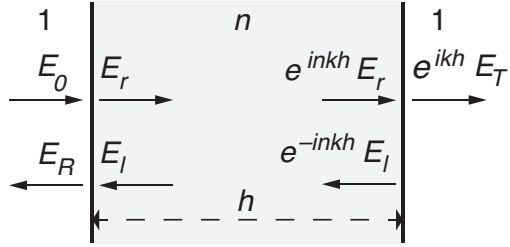}
\caption{Light propagation through a slab of matter: Schematic representation of the electric fields at the surfaces of a slab of thickness $h$ and refractive index $n$. }
\label{SLAB}
\end{figure}

We consider light propagating perpendicularly through a slab of thickness $h$. The light comes in from, and exits to, vacuum with the refractive index 1. We denote the refractive index of the medium (generally complex) by $n$. There are reflected and transmitted waves at the surfaces of the slab, which may be matched in the standard way \cite{JAC99,BOR99}; refer to Fig.~\ref{SLAB}. This leads to the relation between the amplitudes of the incoming and transmitted waves
\beq
\frac{E_T}{E_0}=\frac{2ne^{-ikh}}{2n\cos nkh-i(n^2+1)\sin nkh}\,.
\label{ET}
\eeq
To complete this exercise we note that, according to the local-field corrections the effective electric field inside the sample is \cite{JAC99,BOR99} $E_e = E + P/(3\epsilon_0)$, the polarization is $P=\epsilon_0\chi E$, where $\chi$ is the susceptibility, and also $P=\rho\alpha E_e$, where $\rho$ is the atom density and $\alpha$ is the polarizability of an atom, given by
\beq
\alpha = -\frac{\zeta}{\Delta+i\gamma}, \quad  \zeta=\frac{\dip^2}{\hbar},
\eeq
where $\dip$ is the dipole moment matrix element, $\Delta=\omega-\omega_{{}\,0}$ denotes the detuning of the light frequency $\omega$ from the atomic resonance at $\omega_{{}\,0}$, and $\gamma=\dip^2k^3/6\pi\hbar\epsilon_0$ is the HWHM linewidth of the transition. We then have
\beq
\chi=n^2 - 1=\frac{\zeta\rho}{\Delta-\Delta_{\rm LL}+i\gamma}, \quad  \Delta_{\rm LL}=-\frac{\zeta \rho}{3\epsilon_0} = -\frac{2\pi\gamma\rho}{ k^{3}}\,,
\label{NSQR}
\eeq
where $\Delta_{\rm LL}$ is the Lorentz-Lorenz (LL) shift of the resonance. The power transmission coefficient $T=|E_T/E_0|^2$ and the optical thickness $D=-\ln T$ are easily obtained.

\section{Numerical simulations}
\label{app:num}

In the near-monochromatic case the theory is formulated for the positive-frequency components of electric field, dipole moment, and so on, oscillating at the laser frequency~$\omega$.  Given an atom at position $\bor_i$ in the electric field $\bE(\bor)$, it develops a dipole moment $\bd = \alpha\bE(\bor_i)$, which in turn radiates the dipolar field $\bE_D(\bor) = {\G}(\bor-\bor_i)\bd$. Here $\G(\bor-\bor')$ is the dipole field propagator \cite{JAC99} such that $\G(\bor-\bor') \bd$ is the electric field at $\bor$ from a dipole $\bd$ at $\bor'$. Next take a collection of $N$ atoms at positions $\bor_i$. The field at the position of the $i^{\rm th}$ atom is the sum of the incoming field $\bE_0$ and the dipolar fields from the other atoms,
\beq
\bE(\bor_i)  = \bE_0(\bor_i) + \alpha\sum_{j\ne i} {\G}(\bor_i-\bor_j)\bE(\bor_j)\,.
\eeq
This is a set of linear equations from which one may solve the fields at the positions of the atoms $\bE(\bor_i)$. The total electric field everywhere is then obtained exactly from
\beq
\bE(\bor)  = \bE_0(\bor) + \alpha\sum_{j} {\G}(\bor-\bor_j)\bE(\bor_j)\,.
\label{TOTALFIELD}
\eeq
The total magnetic field is found likewise by adding up the magnetic components of the incoming light and of the dipolar radiation. To model a homogeneously broadened gas, one needs to generate sets of random atomic positions compatible with the geometry of the sample, and average the results over these sets \cite{JAV99LQD}.

The numerical effort starts getting substantial by the time thousands of atoms are involved, so we simulate fairly small circular disks with radius $R$ and area $A=\pi R^2$, and thickness $h$. For maximum compatibility with this geometry, we use circularly polarized incoming light. The convergence to the infinite-slab limit with increasing area of the disk $A$ can be quite poor, and to verify it can be expensive in computer time. We offer a few order-of-magnitude estimates of the truncation error due to the finite area based on spot checks. In all cases we discuss here the statistical error due to the finite number of  samples of atomic positions is at most comparable to, and occasionally orders of magnitude smaller than, the truncation error.

Light is actually not absorbed in a sample such as ours, in which all scattering is elastic. The reduced transmission that we call absorption is because of destructive interference of the incoming light and the light radiated by the dipoles, and the energy that has been removed from the transmitted light simply gets sent in (mostly) other directions. We have studied these assertions in a few cases numerically, and even analytically, in the following way. First, we find the total electric and magnetic fields along the lines of Eq.~(\ref{TOTALFIELD}). Next, we compute the corresponding Poynting vector $\bf S$. Finally, we integrate the normal component $S_\perp$ over a surface, typically sphere, that encloses the atoms. The result was always zero, to within numerical accuracy. This means that in steady state there is no net loss or gain in the energy of the electromagnetic field as a result of the presence of the atoms.

However, the primary outcome of our simulations is the optical thickness $D$ obtained not from Eq.~(\ref{TOTALFIELD}), but by using a method adapted from~\cite{CHO12}. We consider the response of the atomic sample to a plane wave of light propagating in the $z$ direction. For a disk of radius $R$, the forward-scattered light amplitude from a dipole at ${\bf r}_j$ is computed as
\beq
\bE_j({\bf r}) = \frac{ik}{2\pi\epsilon_0 R^2} [\bd({\bf r}_j)-\hat{\bf e}_z \cdot\bd({\bf r}_j)\,\hat{\bf e}_z] e^{ik(z-z_j)}\,.
\label{TRLIGHT}
\eeq
This may  be derived, for instance, by assuming that the dipole is at a random position in a disk of radius $R$, averaging the dipolar field at a distance $\gg k^{-1}$ on the axis of the disk, and taking the limit $R\rightarrow\infty$. The transmitted field is computed by summing the incoming field and the forward scattered fields from all dipoles, which immediately leads to power transmission coefficient and optical thickness.

In elementary test cases this method has proven surprisingly accurate. For instance, knowing the total scattering cross section of an atomic dipole $\sigma(\Delta)$ one can estimate the fraction of radiation scattered away from the forward direction when one atom resides in a disk of radius $R$ as $\sigma(\Delta)/\pi R^2$. On resonance [$\sigma(0)=6\pi k^{-2}$]  the prescription of~(\ref{TRLIGHT}) gives an absorption coefficient ${\cal A}=1-T$ for a single atom that deviates from this ratio by about $2\%$ when the area of the disk is $A=256\,k^{-2}$, and the deviation decreases inversely proportionally to the area of the disk. In the limit of a dilute sample the MFT results are also reproduced as expected, the most conspicuous difference being the density shift of the resonance. This holds true even if the layer of the dilute gas is sufficiently thick that the optical depth is comparable to the ones seen in the simulation result in Fig.~\ref{SLABTRANSMISSION}.

One might think that we could have computed the transmitted intensity without resorting to any approximation by simply having the incoming and the scattered light interfere. However, we have never got this to work satisfactorily. Our atomic samples are quite small, and the concomitant diffraction creates a mode matching problem for interference with the incoming light.  For finite-size disks this would be an issue even in standard optics. The virtue of the forward-scattering approximation is that we can get close to the results expected for an infinite slab with our relatively small atomic samples. Unfortunately, the forward-scattering approximation also removes a potentially important piece of the physics that follows from the discreteness of the radiators, spatial fluctuations such as speckle in the dipolar radiation.

To separate the questions of mode matching and fluctuations we study the reradiated dipolar field only, the sum on the right-hand side of Eq.~(\ref{TOTALFIELD}). We imagine placing a detector that intercepts the transmitted light, a probe disk of the same radius $R$ as the simulation sample, at the distance $kR^2/2$ downstream from the atomic sample. This distance is analogous to the Rayleigh range, approximating the point where the light radiated by the atomic sample starts breaking up from the near-field form of a beam to the far-field form of a cone with a constant opening angle. We integrate the component of the electric field with the same polarization as the incoming beam (which would thus be capable of interference with the incoming field) over the probe disk, and study fluctuations of the integrated field over the atomic samples. If the wavefronts of the incoming light and the reradiated light were perfectly matched, they could cancel completely to give zero transmission. However, this cannot happen in the presence of the fluctuations. For the kind of parameters as in Fig.~\ref{SLABTRANSMISSION} and on resonance, the fluctuations are about 6\%. This would necessarily leave a fluctuating residual field corresponding to the fraction $(6/100)^2$ of the incoming light intensity. The fluctuations therefore limit the maximum optical thickness to about $D\sim-\ln (6/100)^2 \sim 6$. Since we are well below this value in the simulations, we believe that in our examples the fluctuating component in the transmitted light is not a significant feature of the physics.

When the thickness of the disk of Fig.~\ref{SLABTRANSMISSION} is doubled from $h=k^{-1}$ to $h=2\,k^{-1}$ while keeping all other parameters unchanged, the optical thickness in the simulation results also approximately doubles; the maximum becomes $D\simeq5$. The effect is the same as stacking two independent disks with thickness $h=k^{-1}$. We take this to indicate that the thickness in Fig.~\ref{SLABTRANSMISSION} is already sufficient for bulk behavior, just as  $h=k^{-1}$ was in the simulations of Fig.~\ref{SHIFTS}.

\section*{Acknowledgments}
We acknowledge support from NSF, Grant Nos. PHY-0967644 and PHY-1401151, and EPSRC. Most of the computations were done on Open Science Grid, VO Gluex, and on the University of Southampton Iridis 4 High Performance Computing facility.

\end{document}